# Coherent population oscillations with nitrogen-vacancy color centers in diamond

*Authors:*
M. Mrozek[1], A. Wojciechowski[1], D.S. Rudnicki[1], J. Zachorowski[1], P. Kehayias[2], D. Budker[2,3,4], and W. Gawlik[1]

Affiliation:

[1] Institute of Physics, Jagiellonian University, Łojasiewicza 11, 30-348 Kraków, Poland,

[2] Department of Physics, University of California at Berkeley, Berkeley, California 94720, USA,

[3] Helmholtz Institute Mainz, Johannes Gutenberg University, 55099 Mainz, Germany

[4] Nuclear Science Division, Lawrence Berkeley National Laboratory, Berkeley, California 94720, USA

*Abstract*

We present results of our research on two-field (two-frequency) microwave spectroscopy in nitrogen-vacancy (NV⁻) color centers in a diamond. Both fields are tuned to transitions between the spin sublevels of the NV⁻ ensemble in the $^3A_2$ ground state (one field has a fixed frequency while the second one is scanned). Particular attention is focused on the case where two microwaves fields drive the same transition between two NV⁻ ground state sublevels ($m_s=0 \leftrightarrow m_s=+1$). In this case, the observed spectra exhibit a complex narrow structure composed of three Lorentzian resonances positioned at the pump-field frequency. The resonance widths and amplitudes depend on the lifetimes of the levels involved in the transition. We attribute the spectra to coherent population oscillations induced by the two nearly degenerate microwave fields, which we have also observed in real time. The observations agree well with a theoretical model and can be useful for investigation of the NV relaxation mechanisms.

## 1. INTRODUCTION

Nitrogen-vacancy color centers are point defects in the diamond lattice, which consist of a nearest neighbor pair of a substitutional nitrogen atom and a lattice vacancy. The negatively charged NV⁻ centers are used in many areas, e.g., as fluorescent markers for biological systems, for quantum information processing or for sensing electric and magnetic fields [1–7]. One of the techniques used to probe NV⁻ centers is optically detected magnetic resonance (ODMR), where one detects the fluorescence intensity changes which correspond to the ground-state population changes induced by resonant microwave (MW) fields [8, 9]. The excitation, spin polarization, and interrogation of the ground-state spin are often done with green laser light.

In a recent work [10], it was shown that a strong MW field can burn a hole in the NV⁻ ODMR spectrum that can then be probed with a weak MW field, similarly to the case of nonlinear Doppler-free laser spectroscopy of gas samples [11,12], or spectral hole burning in solids [13-16]. In the case of crystals, NV diamond in particular, the different local strain and magnetic fields are the sources of inhomogeneous broadening, rather than Doppler effect. The MW hole-burning technique removes the inhomogeneous broadening and was shown to be useful for magnetic-field-insensitive thermometry at room temperature [10]. The experiment studied the case where the two MW fields were tuned to two *distinct* transitions of the V configuration in the NV⁻ ground state manifold, i.e., $m_s = 0 \leftrightarrow m_s = -1$ and $m_s = 0 \leftrightarrow m_s = +1$, where $m_s$ is the spin projection on the axis of the NV⁻ center (Fig. 1).

In the present work we also observed two-field nonlinear and interference effects within the NV⁻ ground state, but with two phase-coherent MW fields acting on *the same* transition, $m_s = 0 \leftrightarrow m_s = +1$.

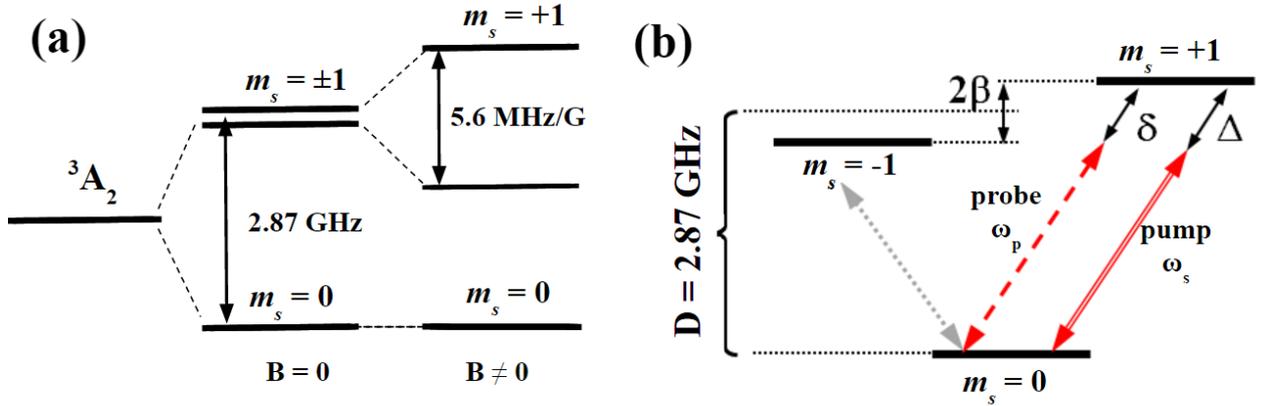

Fig.1. (a) The energy level structure of the NV⁻ ground-state $^3A_2$ without (B=0) and with (B≠0) magnetic splitting. b) Transitions induced by the pump (double solid arrow) and probe (broken arrow) microwave fields. The pump acts on the $m_s = 0 \leftrightarrow m_s = +1$ transition, while the probe can be tuned to either of the $m_s = 0 \leftrightarrow m_s = \pm 1$ transitions. The pump and probe field detunings from resonance are labelled as $\Delta$ and $\delta$, respectively.

The resonances observed at the same transition differ from those studied previously with two distinct transitions: they have larger amplitude and exhibit complex multi-Lorentzian shapes. We attribute the resonances occurring on the same transition to the coherent population oscillations (CPO) resulting from interference (beating) between two coherent electromagnetic fields [17, 18]. The notion of CPO is widely used in the atomic physics/quantum optics community but is essentially unknown in the EPR/ODMR community. With this paper we want to show that the CPO ideas are applicable and useful also for EPR/ODMR field.

The beating of two coherent waves leads to a temporal modulation of the population difference between the upper and lower states, described in literature as the coherent population oscillations. The population oscillations combine with the incident waves and enhance them or attenuate, depending on relative phases of the population and wave beating. Because of the population inertia, the effect occurs only for beat frequencies within the range determined by the population relaxation times. As demonstrated below, the CPO effect changes drastically the shape of saturation hole within a very narrow spectral range, on the order of the population relaxation rates.

We note that closely related effect occurs when one modulated wave is used. The sidebands created by such modulation act similarly to two independent waves and also cause population modulation. There is a difference, though, between the amplitude or power modulation of a single beam and the two waves: in the former case the modulation does not affect the carrier wave, in the later the net amplitude is sensitive to the modulation phase. The beat-note of the resulting net power is identical in both scenarios. However, some differences can be expected when the dynamics of the system is governed by the amplitude rather than the power of the field. Appendix 1 presents detailed comparison of the case of two coherent waves with the amplitude modulation of a single field.

While the basic features of the hole burning with distinct transitions are satisfactorily explained by the population saturation, the case of the same transition requires accounting for the CPO interference effects. In particular, the simple hole-burning model [11] fails to predict the appearance of saturation dips if the transition under consideration is only homogenously broadened [19]. Such dips are, however, described theoretically [19-22] and detected experimentally in a number of systems, including *p*-type semiconductors [23], semiconductor quantum wells [24], ruby crystals [25], Sm+2:CaF2 crystals [23], erbium-doped optical fibers [27], and atomic vapors in room [28-30] and sub-mK temperature [31].

The phenomenon of CPO plays an important role in nonlinear spectroscopy [11, 20-22,32] and wave-mixing experiments [33-36]. Recently, it attracted much attention because of its potential for light storage [28-31, 37,38]. So far, however, the effect has not been directly observed with diamond. Very interesting indirect observation was recently presented by Golter *et al.* who studied CPO-related interference effects between modulated optical and optomechanical excitations [39].

To interpret our observations of CPO, we adopted the laser spectroscopy approach of Baklanov and Chebotayev [20,21] and applied it to our experiment with ensemble NV diamond and two continuous-wave (cw) MW fields. We reproduced qualitatively the experimental observations, and particularly the difference between the resonances on distinct and the same transitions. By controlling the phases of the two MW fields we were able to directly observe the CPOs as a function of time (in real time), while by changing the frequency difference between the pump and probe fields we resolved complex internal structure of the hole-burning resonance, absent for the case of distinct transitions. As demonstrated below, analysis of this structure should enable determination of the relaxation rates of the spin states in the ground state of the NV diamond sample.

The paper is organized as follows: Sec. 2 describes the experimental conditions, Sec. 3 presents our results, and Sec. 4 explains the modeling of the observed effects by an analytical approach within a steady-state approximation. Section 5 is devoted to the observation and time-dependent analysis of population oscillations. Conclusions are presented in Sec. 6.

## 2. EXPERIMENTAL CONDITIONS

The goal of the experiment was to study the simultaneous interaction of two microwave fields with the NV$^-$ color centers in a diamond at room temperature. The MW fields were tuned to the transitions depicted in Fig. 1(b).

A schematic diagram of the experimental setup is shown in Fig. 2. In our measurements we used a diamond sample with initial nitrogen density below 200 ppm manufactured by Element Six by high-temperature high-pressure growth, 3×3×0.3 mm$^3$ in size, and cut along the {100} surface. The sample was irradiated with an electron beam (14 MeV, fluence 1.5×10$^{18}$ cm$^{-2}$) and annealed for two and a half hours at 750 ºC. The resulting NV$^-$ concentration is about 20 ppm.

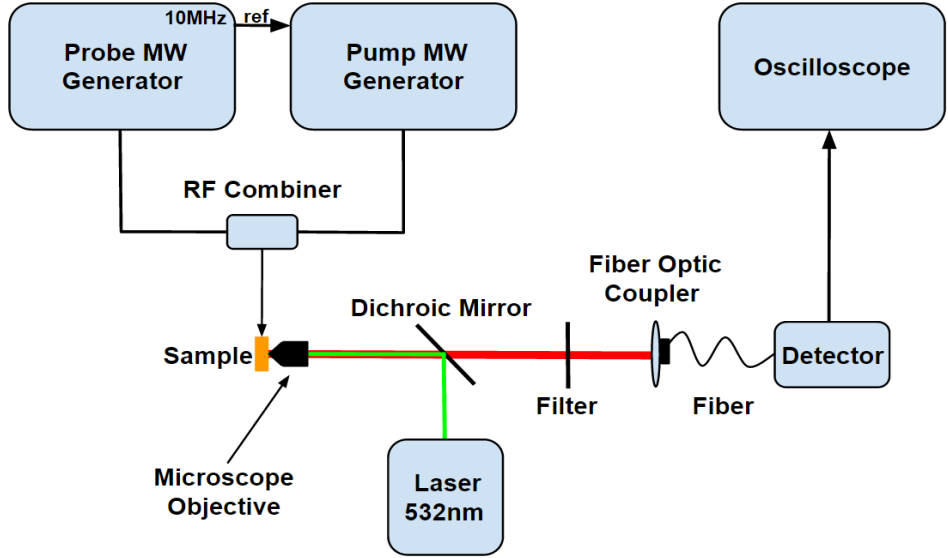

Fig.2. The experimental setup for hole burning experiments.

The optical part of the setup utilizes a confocal arrangement. The sample is optically excited using 532 nm laser light (5 mW power), and the fluorescence is recorded in the 600–800 nm range through the same microscope objective (Olympus 40×, numerical aperture 0.75) and an avalanche photodiode (50 MHz bandwidth).

The MW fields, pump (fixed frequency $\omega_s$ tuned close to the $m_s = 0 \leftrightarrow m_s = +1$ central frequency) and probe (variable frequency $\omega_p$, scanned linearly in time) were created using two generators, combined, and connected to the microstrip structure [40] attached to the sample. The field detunings from exact resonances were $\Delta = \omega_s - \omega_{+10}$ and $\delta = \omega_p - \omega_{\pm 10}$, where $\omega_{\pm 10}$ is the central frequency of the $m_s = 0 \leftrightarrow m_s = \pm 1$ transition. The effect of two cw MW fields on the NV⁻ ground-state populations was monitored by optical fluorescence (the ODMR technique). The RF power was calibrated by measuring the Rabi oscillation frequency. We used Rabi frequencies to characterize the MW fields as they allow a better comparison between the spectra recorded under different conditions, like different distances of the spot on the sample from the MW antenna. Typical values of the on-resonance Rabi frequencies were 2π×1.6 MHz for the pump and 2π×1.2 MHz for the probe, unless specified otherwise. The scan of the probe frequency had a staircase shape, with a fixed number of steps (typically 16000) and the overall time of scan of 100 ms. The phase of the probe MW field was preserved during the jumps between discrete frequency steps of the frequency synthesizer.

In a magnetic field $B=28$ G, aligned along the [111] diamond axis, the linearly polarized MW fields were driving well-resolved transitions between the $m_S = 0 \leftrightarrow m_S = \pm 1$ ground states (Fig. 1). The Zeeman splitting for the NV⁻ centers oriented along this direction equals $2\beta = 2g\mu_B B$ where g is the Landé factor, $\mu_B$ is the Bohr magneton, and $B$ is the magnetic field, while for the other three alignments it amounts to $2\beta = 2g\mu_B B|\cos(109.5°)|$. In this work we focus on the case when both MW fields are tuned to the same transition between the $m_s = 0 \leftrightarrow m_s = +1$ ground states. Consequently, the studied transition is essentially between the states of a two-level system. An additional difference between the current experiment and the one in Ref. [10] was that one of the generators provided the 10 MHz frequency reference for the second one. Such synchronization allowed observation of the interference effects for small (sub-kHz) detunings of the two fields with no detectable phase drift of the two generators over whole scans.

## 3. RESULTS

Figure 3(a) shows the ODMR spectra recorded with the magnetic field of 28 G oriented along the [111] crystallographic direction. The black curve presents the regular ODMR spectrum recorded as a function of $\omega_p$ without the pump field. It consists of four Gaussian-like resonances corresponding to inhomogeneously broadened transitions. The inhomogenous width (~10 MHz) is comparable to that observed in the cw case in Ref. [10] and results from various contributions, like local magnetic field and/or strain inhomogeneity, and unresolved hyperfine structure with a possible power broadening. For the red curve the pump field was applied with the frequency 2948.5 MHz, which matched the $m_s = 0 \leftrightarrow m_s = +1$ transition frequency for the NV$^-$ centers aligned along the [111] crystal direction. As seen in the figure, the presence of the pump field resulted in burning of two holes: one at the $m_s = 0 \leftrightarrow m_s = +1$ transition, and another one around 2788 MHz, which corresponds to the frequency of the $m_s = 0 \leftrightarrow m_s = -1$ transition for the same crystallographic orientation. No effect of the pump field is seen in the other ODMR components corresponding to three other possible alignments of the NV$^-$ centers. This fact reflects the sensitivity of MW saturation/hole burning to the alignments of the NV$^-$ subensemble in the diamond sample.

The shapes of the holes burned in the two transitions ($m_s = 0 \leftrightarrow m_s = \pm1$) differ considerably: for the case of pumping and probing on the same MW transition, the resonance is sharper and deeper than that appearing at the different transition and clearly non-Lorentzian, i.e., exhibiting internal structure [Fig. 3(c)]. Moreover, positions of the two holes are correlated such that when the pump beam (and one hole) is at $\omega_{+10}-\Delta$ [Fig.3(c)], the second hole appears around $\omega_{-10}+\Delta$ [Fig.3(b)]. This observation supports the conclusion of Ref. [10] that the inhomogenous broadening of the observed ODMR lines is primarily caused by the local magnetic field, rather than strain variations [41]. Hereinafter we refer to the situation when the pump and probe fields interact with the same transition ($m_s = 0 \leftrightarrow m_s = +1$) as to the +/+ case, while that with the pump at the $m_s = 0 \leftrightarrow m_s = +1$ and the probe at $m_s = 0 \leftrightarrow m_s = -1$, as the +/− case. In this work we concentrate on the +/+ case.

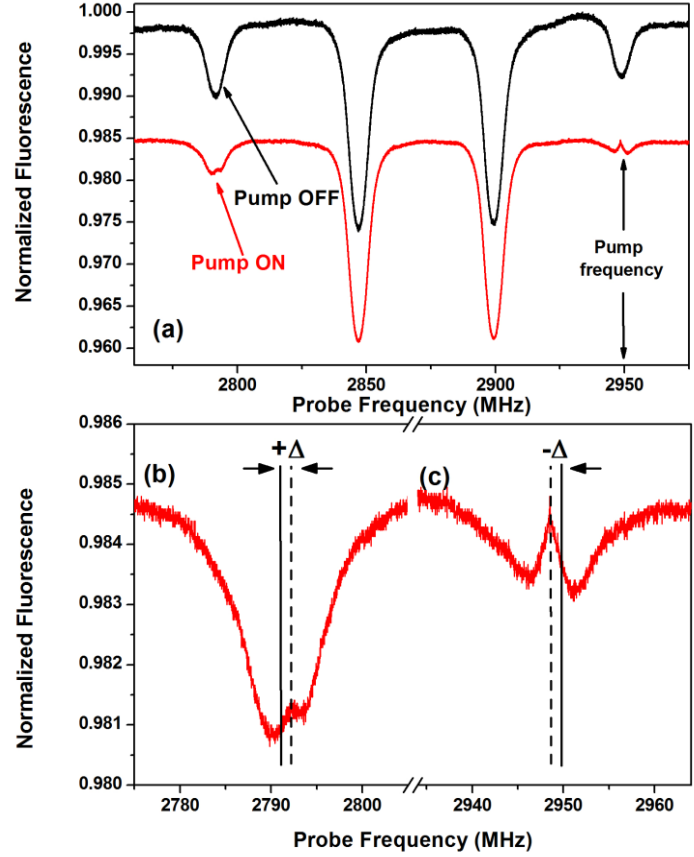

Fig.3. (a) ODMR spectrum without (black) and with (red) pump MW field in the static magnetic field of 28 G along the [111] direction. In the rightmost line the pump and probe act on the same transition (the +/+ case) and the effect of hole burning is well visible; The two central lines correspond to other orientations of the NV⁻ center; in the leftmost line the pump and probe act on two different transitions of the V structure (the +/- case). (b, c) Zoom-in on the ODMR lines with burned holes in the +/- and +/+ case, respectively. Solid lines mark the centers of the inhomogeneously broadened lines.

Figure 4 presents the dependence of the shape of the holes burned in the inhomogeneously broadened ODMR lines on the pump power [Fig. 4(a)] and frequency [Fig. 4(b)]. For the sake of comparison, we present side by side the +/+ normalized spectra along with the corresponding +/- ones. In both cases MW pump field causes an increase in the fluorescence level, i.e. burns a hole in the population of the $m_s = 0$ state. In the +/+ case, a central narrow (width on the order of 10 kHz) peak on top of the wide (~1 MHz) pedestal is also visible [Fig. 4(a) right spectra]. The position of these features is determined by the pump frequency $\omega_s$ relative to the frequency of the transition $m_s = 0 \leftrightarrow m_s = +1$. For high pump powers the fluorescence level approaches the off-resonant values (saturates), the pedestal broadens, and the central peak vanishes. We attribute the observed narrow structures to the CPO effect, and discuss the complex hole lineshape in more detail in the next section. For the +/- situation [Fig. 4(a) left spectra] only the wide pedestal is visible, and the fluorescence level is generally lower, i.e., the ODMR signal has higher contrast than for the +/+ case. With increasing pump power the inhomogeneously broadened line flattens due to the increase of the hole amplitude and its power broadening. In Fig. 4 (b) the spectra obtained for different frequencies of the pump field are presented. In the +/+ case, when $\omega_s$ is tuned to the side of the inhomogeneously broadened profile, the burned holes appear as skewed profiles. The central narrow structure, however, preserves its shape and appears always at the pump frequency. The maximum fluorescence level remains constant independently of pump detuning.

More regular hole shapes are observed in the +/- case, however their low contrast hinders accurate quantification of hole positions for not-too-strong MW powers. This can be seen, e.g., on the green curve in Fig. 4 (b), where the +/- profile shows an apparent shift towards higher frequencies, due to the line-pulling effect of the hole burned on the lower frequency side of the inhomogeneously broadened line.

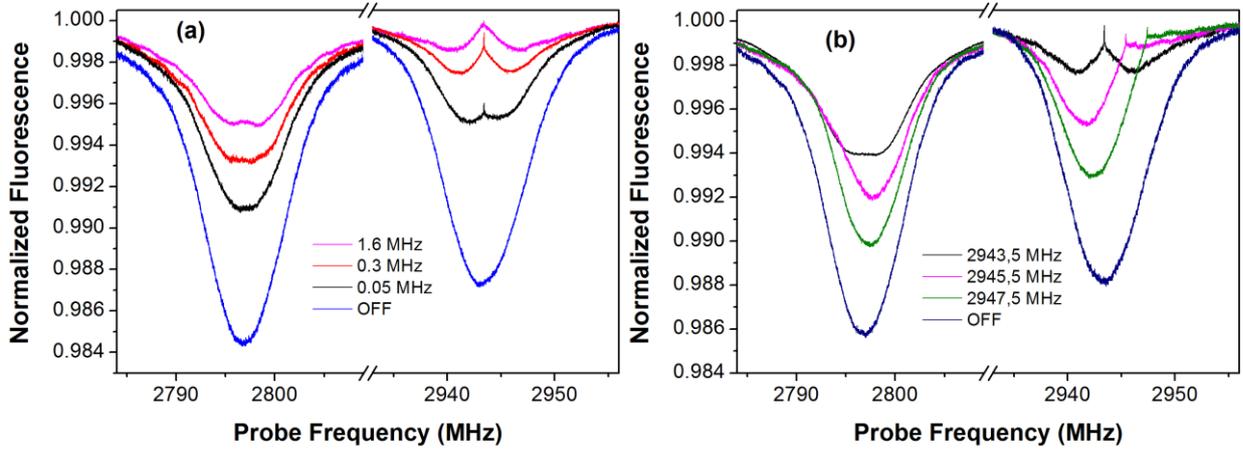

Fig.4. Comparison of the spectral hole-burning spectra recorded for: (a) probe power of 1.2 MHz and different pump powers (in Rabi-frequency units), and (b) probe frequencies (for pump power of $2\pi \times 1.0$ MHz and probe power $2\pi \times 0.5$ MHz). The left-hand side resonances in (a) and (b) represent the +/- case, while the right-hand side ones correspond to the +/+ case. The curves have been normalized individually (to the off-resonance level) to show the relative fluorescence changes.

The same qualitative dependences (frequency shifts and power broadening) we have observed also with other samples although the quantitative details vary from sample to sample.

## 4. STEADY-STATE MODELLING

In order to model the ODMR resonances arising from the interaction of the two MW fields in the +/+ case, we employ the method developed for laser spectroscopy with two-level atomic systems by Baklanov and Chebotaev [20,30] and Sargent *et al.* [22]. This approach has been extended by Boyd and Mukamel [19] by taking into account interference of all possible time orderings of the interacting fields. For the +/+ case with well-resolved ODMR transitions, the two-level system composed of $m_s = 0$ and $m_s = +1$ is sufficient to describe the basic mechanism of hole burning and probing by two independent fields.

Evolution of a two-level system perturbed by two fields of frequencies $\omega_s$ and $\omega_p$ may be characterized with the help of a density matrix formalism, by the matrix element between states $m_s = 0$ and $m_s = +1$, i.e. the coherence $\rho_{01}$. Coherence $\rho_{01}$ contains time dependent contributions oscillating not only at the field frequencies $\omega_s$ and $\omega_p$ but also at the combination frequencies $2\omega_s-\omega_p$, and $2\omega_p-\omega_s$ and their harmonics due to nonlinear wave mixing. The number of the harmonics increases with the power of the interacting fields as more waves are generated and themselves contribute to the wave mixing. In the optical domain these contributions have been interpreted as various nonlinear phenomena such as wave-mixing or coherent population oscillations [19, 21, 32, 33].
In this section we demonstrate that analogous wave coupling is possible in solid-state samples perturbed by MW fields. Moreover, the difference between the optical and MW frequency domains enables observation of interesting dynamics of interfering MW fields, which in the optical range is typically observed only as a time-averaged quasi-stationary response.

Following the formalism of Ref. [20,21], applicable for a weak probe, the leading contribution to $\rho_{01}$ contains three time-dependent terms:

$$\rho_{01}(t) = R e^{i\omega_s t} + i\Omega_p A e^{i\omega_p t} - i\Omega_p B^* e^{i(2\omega_s - \omega_p)t}, \qquad (1)$$

where *R*, *A*, and *B* denote slowly-varying amplitudes of the components of the coherence oscillating with frequencies $\omega_s$, $\omega_p$, and $2\omega_s$-$\omega_p$, respectively. Quantities *R, A, B* are nonlinear functions of the pump field Rabi frequency $\Omega_S$. The effect of the probe is accounted for by the two terms proportional to the Rabi frequency $\Omega_p$. In the model it is assumed that $\Omega_p \ll \Omega_S$ which neglects all higher-order terms in $\Omega_p$ in Eq.(1). We assume here that optical excitation establishes initial populations of the $m_s$-states by pumping most of NVs to the $m_s = 0$ state and enables detection of these populations but has no other effect on the system dynamics.

Since the MW fields couple non-diagonal elements of the density matrix (contributions *R, A, B*) with the diagonal ones (populations), the oscillations of the coherence components in Eq. (1) result in the related oscillations of the $m_s$ state populations. The effect appears most strongly when the initial fields $\omega_s$, $\omega_p$, and the additional one $2\omega_s$-$\omega_p$ beat together, i.e. at the center of the +/+ hole at $\omega_s = \omega_p$. This phenomenon is described in Ref. [20-22] as the *modulation* or *coherence* effects and as *coherent population oscillations* in Ref. [19, 33].

$$\rho_{11} - \rho_{00} = (\rho_{11} - \rho_{00})^{dc} + (\rho_{11} - \rho_{00})^{(\Delta_\omega)} e^{i\Delta_\omega t} + (\rho_{11} - \rho_{00})^{(-\Delta_\omega)} e^{-i\Delta_\omega t}, \qquad (2)$$

where the subsequent terms represent, respectively, the stationary population difference and the contributions oscillating as $\pm \Delta_\omega$, explicitly responsible for CPO.

The formalism of Refs. [20,21] needs to be considered as the first-order approximation to a more complete picture. In particular, if the probe field becomes comparably strong as the pump, Eq.(1) would include also terms oscillating as $2\omega_p$-$\omega_s$ and at other combination frequencies, i.e. at higher harmonics of $\Delta_\omega = \omega_p$-$\omega_s$. Still, even that simplified approach reproduces the main features of the CPO effect and allows us to interpret the experimental results.

For the +/- case, the relevant frequencies differ strongly, hence the beating effect is negligible because the beat frequency is usually faster than the NV$^-$ reaction time characterized by the relaxation rates of the involved spin states. Consequently, the MW absorption spectrum $\kappa'(\omega_p)$, for $\omega_p$ scanned around the $\omega_{-10}$ frequency, recorded when the pump field frequency $\omega_S$ is tuned close to the $\omega_{+10}$ resonance frequency, is mainly characterized by hole-burning in the population of the $m_S=0$ state, with the Lorentzian profile of the hole:

$$\frac{\kappa'(\omega_p)}{\kappa} = e^{-\left(\frac{\omega_{-10}-\omega_p}{W}\right)^2} \left(1 - \frac{G}{2}\frac{4\Gamma^2}{(\omega_{-10}-\Delta-\omega_p)^2 + 4\Gamma^2}\right), \qquad (3)$$

where: $\kappa'(\omega_p)$ denotes the absorption coefficient of the probe MW field and $\kappa$ is its on-resonance nonsaturated value, *W* is proportional to the inhomogeneous width of the $m_s = 0 \leftrightarrow m_s = -1$ transition, $G = 2\Omega_S^2/(\gamma\Gamma)$ is the saturation parameter associated with the pump-field Rabi frequency $\Omega_S$, $\Gamma$ is the homogenous linewidth, $\gamma_0$ and $\gamma_1$ are the population relaxation rates of state $m_S = 0$ and $m_S = +1$, respectively, and $\gamma = \gamma_0\gamma_1/(\gamma_0+\gamma_1)$. Expression (3) reflects the inhomogeneously broadened absorption profile with a saturation dip well known from laser spectroscopy. The dip (hole) depth is determined by the value of $G$ while $\Gamma$ sets its width. Neither optical nor MW power broadening is considered in this low-field model. The rates $\gamma_0$, $\gamma_1$, and $\Gamma$ are directly related to the standard $1/T_1$, $1/T_2$ longitudinal and transverse rates: $1/T_1 = (\gamma_0+ \gamma_1)/2$ and $1/T_2 = \Gamma$. In a general case, when coherence decay is enhanced by some dephasing, the transverse relaxation is faster than that resulting from the relaxation of populations: $\Gamma=(\gamma_0+\gamma_1)/2+\Gamma_d$ where $\Gamma_d$ is the dephasing rate [42].

For the +/+ case, the above discussed beating and the resulting population oscillations become more important. The analysis of Refs. [20-22] based on the steady-state approximation yields for this case an analytical expression, which for not-too-strong pump field ($G<<1$) can be cast into an approximate form:

$$\frac{\kappa'(\omega_p)}{\kappa} = e^{-\left(\frac{\omega_{+10}-\omega_p}{W}\right)^2} \left\{1 - \frac{G}{2}\frac{4\Gamma^2}{\Delta_\omega^2+4\Gamma^2}\left[1-\frac{\gamma}{\Gamma}+\left(\frac{\gamma}{\gamma_0}+\frac{\gamma}{2\Gamma}\right)\frac{\gamma_0^2}{\Delta_\omega^2+\gamma_0^2}+\left(\frac{\gamma}{\gamma_1}+\frac{\gamma}{2\Gamma}\right)\frac{\gamma_1^2}{\Delta_\omega^2+\gamma_1^2}\right]\right\}. \qquad (4)$$

In the experiment with NV⁻ centers we do not directly measure the absorption coefficient of the probe field as in Ref. [20]. Instead, we apply optical detection of luminescence induced by the green light, which, on the one hand, creates spin polarization in the ground state, represented by the initial values of populations $n_0^0$ and $n_1^0$, and, on the other hand, is sensitive to the actual $m_s$ state populations and their difference. Consequently, the populations of the $m_s$, and the absorption features of the $\kappa'$ coefficient [Eqs. (3, 4)], transform into the corresponding fluorescence-intensity variations seen as the hole-burning resonances in ODMR spectra.

The main difference between the spectra described by Eqs. (3) and (4) is the shape of the saturation hole burnt by the pump in the absorption spectrum. In the +/+ case the hole is composed of several Lorentzian contributions with relative amplitudes: $1-\gamma/\Gamma$, $\gamma/\gamma_0+\gamma/2\Gamma$, and $\gamma/\gamma_1+\gamma/2\Gamma$ and half-widths: $2\Gamma$, $\gamma_0$, and $\gamma_1$, respectively. If the relaxation rates $\gamma_0$, $\gamma_1$, and $\Gamma$ are comparable, and particularly if there is no dephasing and $\Gamma = (\gamma_0+\gamma_1)/2$, the dips observed in the +/- and +/+ case are not very different. The situation changes significantly, however, when the dephasing becomes significant, $\Gamma >> \gamma_0$, $\gamma_1$. In that case, the superposition of all contributions to Eq. (3), yields a hole, which is deeper and more pointed than that predicted by Eq. (3) for the +/- case. The presented modeling of the ODMR spectra with two MW fields reproduces the observed features, as illustrated by the hole shapes presented in Fig.3(b) and 3(c).

For a detailed comparison, Figure 5 presents the +/+ signals recorded in a wide (a) and narrow (b) range of $\omega_p - \omega_s$, along with the fitted signals (red solid lines). The fitting curve is a sum of a Gaussian background which represents the inhomogeneously broadened ODMR line and three Lorentzian components occurring at $\omega_p = \omega_s$ as predicted by Eq.(4). The fitted lineshape is in agreement with the recorded data, with the broad pedestal, as well as the intermediate and narrow structures. The Lorentzian contributions have the widths (HWHM) of 27(7) kHz, 137(49) kHz, and 2.903(38) MHz. In our modeling they are attributed respectively to the $\gamma_0$ and $\gamma_1$ (the two lower values) and $2\Gamma$ (the highest value) relaxation rates of the NV⁻ population and coherence under the influence of the optical excitation light and MW fields. The signals shown in Fig. 5 were recorded with the steps of 3.125 kHz per 6 μs. As the width of the recorded narrowest Lorentzian could be affected by the scan rate, we have also recorded narrow scan-range spectra of the central structure with higher resolution. These data revealed interesting additional oscillatory structure, which is discussed in detail in the next section.

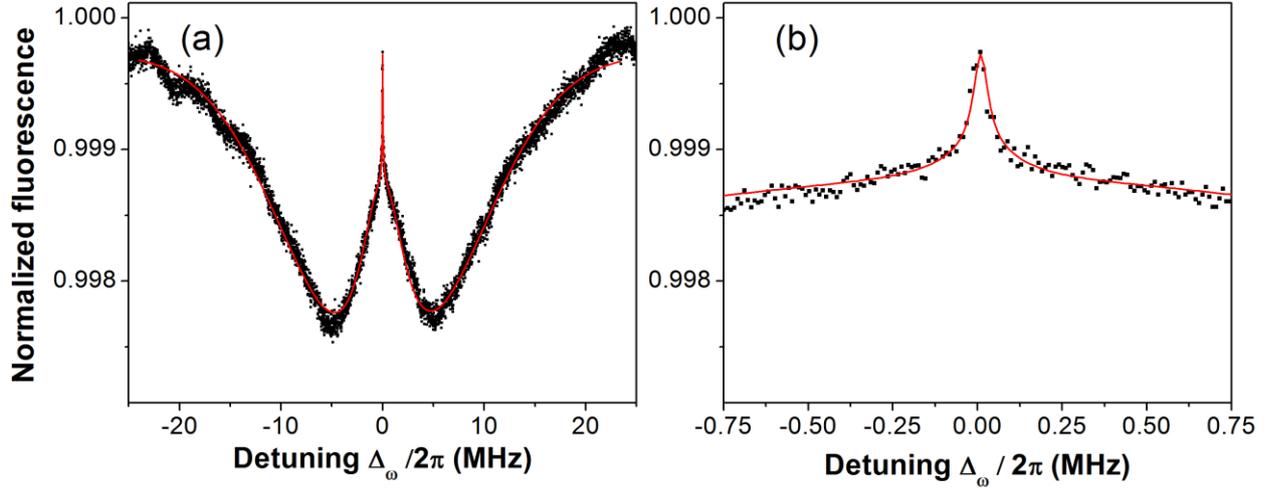

Fig.5. Central part of the ODMR hole burned in the +/+ case with the pump field tuned to the center of the $m_S = 0 \leftrightarrow m_S = +1$ transition (a) and its zoom (b). The fitted curve (solid red) is a sum of Gaussian background dip (HWHM = 10,037(33) MHz) and three peaks centered at $\omega_s$ with widths, respectively: 27(7) kHz, 137(49) kHz, and 2.903(38) MHz).

The individual contributions to Eq. (4) are most easily separated when the relevant relaxation rates $\gamma_0$, $\gamma_1$, and $\Gamma$ differ strongly. When the rates are comparable, i.e. when the individual contributions to the overall resonance have comparable widths, their separation is not easy. In the Appendix we present results of two fitting procedures applied to the data of Fig. 5, involving two and three individual contributions. While raw observation of the resulting lineshape does not allow immediate identification of all three components, the fitting procedure supports interpretation in terms of three contributions discussed above.

## 5. POPULATION OSCILLATIONS

Figure 6 shows the ODMR signal for the +/+ configuration, recorded under the same conditions as in Figs. 3(c) and 5 but with higher spectral resolution. In Fig. 6 (a) the inhomogeneously broadened (~10 MHz linewidth), regular ODMR profile is seen with the complex structure of wide and narrow peaks at $\omega_p=\omega_c$. Successive plots (b – d), with decreasing frequency span, gradually reveal additional oscillatory structure with an increasing amplitude. As seen in Fig. 6(d), the phase of oscillations depends quadratically on the detuning (time), which results from our linear frequency sweep. The oscillatory structure represents the coherent population oscillations mapped onto corresponding oscillations of the fluorescence intensity, as discussed in Sec.4. When two MW fields have sufficiently similar frequencies, their interaction with the sample could be considered as an interaction of a single effective field, with slowly varying amplitude/power due to the wave beating. In such case, the populations follow adiabatically the effective field and result in the observed oscillations of the fluorescence intensity. The fluorescence level at $\Delta_\omega = 0$ in Fig. 6 (c) and (d) reflects the relative MW phase and varies from one scan to another over the whole oscillation amplitude, depending on the initial phase at the beginning of the scan. The asymmetry between the top and bottom envelope profiles and the non-sinusoidal form of the oscillation signal are manifestations of nonlinearity of the MW-NV interaction. The width of the envelope depends on the frequency scan rate and time, thus neither the oscillations, nor the envelope seen in Fig. 6 (c and d) should be considered as direct spectral characteristics of the signal. They are just representations of single realizations of the linear frequency scans. If, however, the frequency of the probe MW field is scanned at a fast rate, the populations cannot follow the effective field and the oscillations become averaged. This behavior is most evident in Fig. 6 (c) as a decrease of the oscillation amplitude.

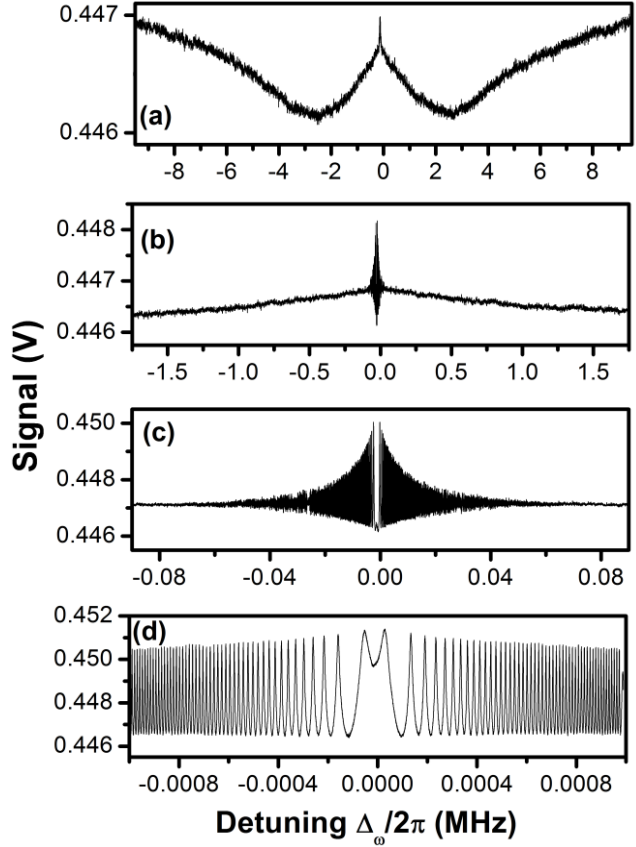

Fig.6. The +/+ ODMR resonance observed with $\omega_S/2\pi = 2948.5$ MHz, i,e. at the central frequency of the $m = 0 \leftrightarrow m_s = +1$ transition and magnetic field of 28 G in the [111] direction. For all plots the scan time was 100 ms and total number of frequency steps was 16000 with the different frequency step size.

Since the fluorescence signal oscillates at the beat-note frequency, a stable spectral response can be seen only if time/phase-averaging is provided, e.g., by averaging scans with different relative phases of the two MW fields. To find the true spectral characteristics we have summed a few thousand recordings taken for different relative phases of the two MW fields, distributed uniformly over $2\pi$. Figure 7 presents the comparison of such a phase-averaged signal with a similar set recorded for a fixed initial phase of the MW fields. The trace of latter exhibits a high-amplitude oscillation with the fluorescence level approaching that without MWs [black curve in Fig. 3a, outside the resonances], since the maximum fluorescence occurs at instances of the destructive interference of the two fields. While the envelope of the single-phase realization depends on the scan rate, the phase-averaged trace is insensitive to the scanning rate and shows a Lorentzian-shape of reduced height and the HWHM width on the order of 1 kHz. Thus, the phase-averaged scans enable us to assign the width of the central feature to the slowest of the relaxation rates ($\gamma_0$ or $\gamma_1$). In contrast to the optical domain and typical laser sources where phase-randomization results from the finite spectral width of the lasers used, the MW sources may preserve the phase for many scans and explicit averaging mechanism has to be applied to extract the spectrum from the frequency scans.

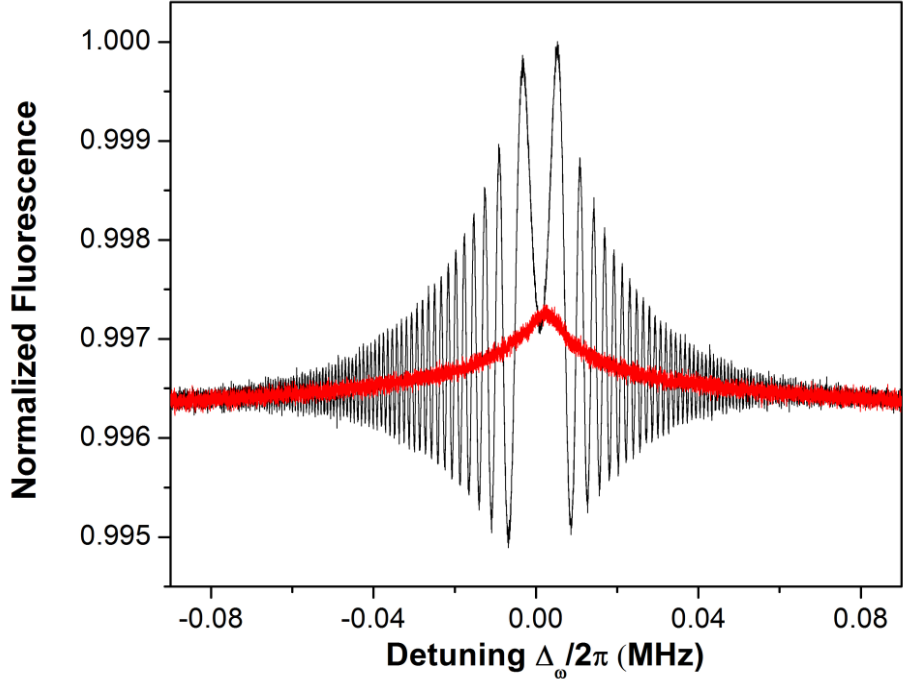

Fig.7. Averaged fluorescence signals as a function of $\Delta_\omega$ with fixed (black curve) and evenly distributed (red curve) initial relative MW phase of each scan. Data were taken with Rabi frequencies ~ $2\pi \times 100$ kHz.

It was shown above that the signatures of population oscillations can be observed in the spectral domain, which requires averaging for a time longer than the characteristic oscillation period or equivalent phase averaging. Apart from this, for the very small detunings $\Delta_\omega$ we were able to observe the population oscillation directly by recording the time-dependent NV$^-$ fluorescence for a fixed value of the detuning. Figure 8 (a) presents the CPO oscillations observed in real time with detuning between the pump and probe MW fields $\Delta_\omega/2\pi = 1$ kHz. The oscillations are non-sinusoidal and contain many harmonics as shown in the Fourier spectrum of the time dependence in Fig. 8 (b). The number of harmonics rapidly increases with the MW power.

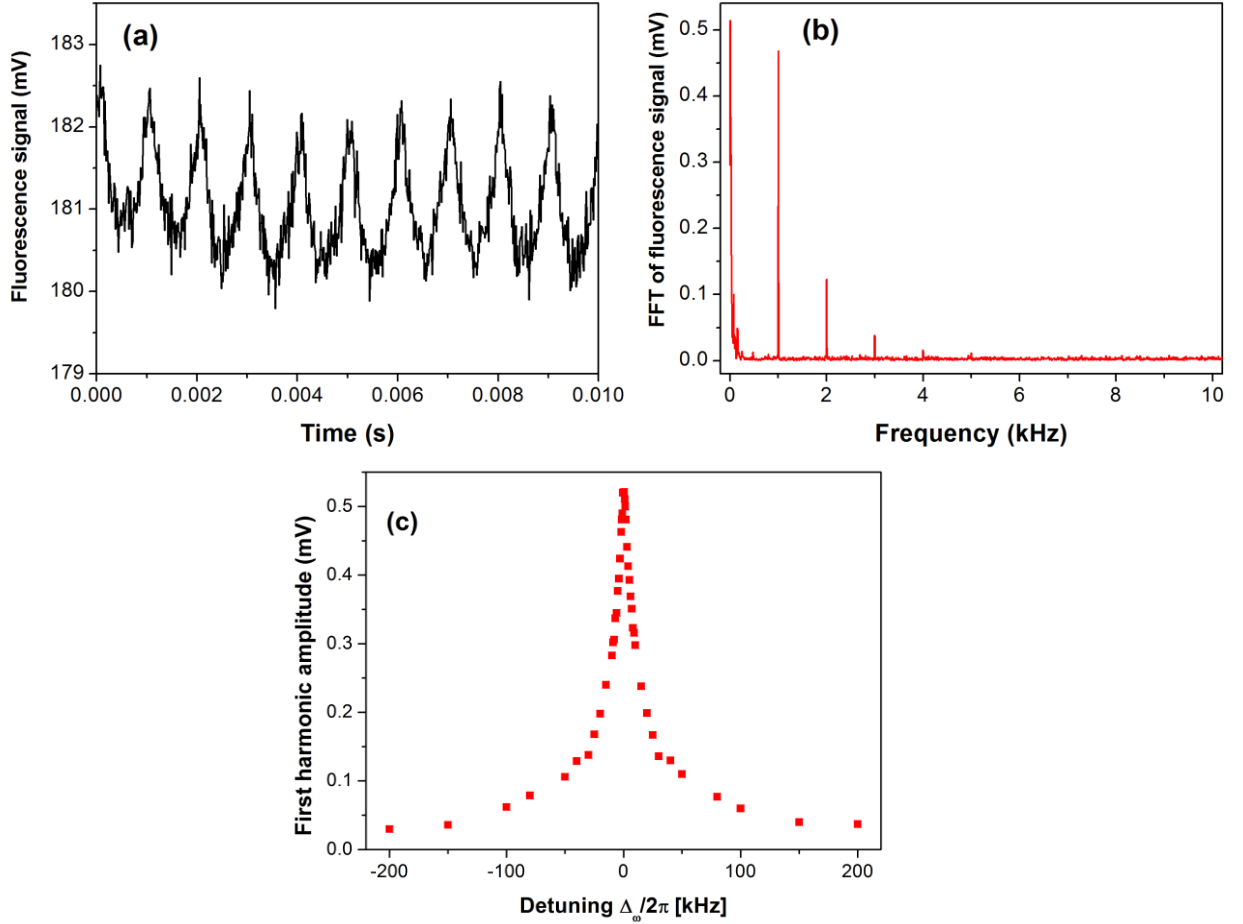

Fig. 8. (a) Time-domain of the ODMR signal ODMR signal for $(\omega_s-\omega_p)/2\pi = 1$ kHz. (b) Fourier transform of the signal shown in (a). (c) Amplitude of the first harmonic peak of the Fourier-transformed signals as a function of the detuning $\Delta_\omega/2\pi$ in the range from -200 to 200 kHz with 1 kHz being the lowest difference measured.

In Figure 8 (c) we demonstrate how the amplitude of CPO oscillations extracted from the FFT of the fluorescence signals depends on detuning $\Delta_\omega$. The amplitude is rapidly reduced when $\Delta_\omega$ increases but the oscillations are still visible even for detunings on the order of 100 kHz. Interestingly, the oscillation envelope [Fig. 8 (c)] resembles the shape of the narrow resonance demonstrated in Fig. 5 (b) for the +/+ case, although Fig. 8 (c) displays only the first harmonic of the beat signal, whereas all harmonics contribute to the cw resonances. Given the observed oscillations occur only when the frequencies of two MWs are close to the specific frequency of the spin transition between the NV spin states, we find that our observations rule out the possibility of a trivial electronic wave beating as an origin of the observed oscillations.

## 6. TIME-DEPENDENT MODELLING

By solving density matrix equations one can calculate not only optical coherence for analysis of spectral dependences such as described in a previous section, but also the populations $\rho_{00}$ and $\rho_{11}$ of our model two-level system and their time dependence. Similarly as in Refs. [39, 43] one may adiabatically eliminate the optical coherence $\rho_{01}$ and arrive at rate equations that describe the time and phase dependence of the populations on time-scales much longer than $1/\Gamma$.

Here, we take a more intuitive approach which gives more insight to the phenomenon of CPO. We take advantage of the fact that for very small frequency difference of ω$_S$ and ω$_p$ the net MW field becomes quasi stationary. Rather that analyzing each of the fields separately, we take into account the effective slowly varying MW field, resulting from the joint action of both MW fields. Equations for the populations $\rho_{11}$ and $\rho_{00}$ read then:

$$\dot{\rho}_{11} = -\gamma_1(\rho_{11} - n_1^0) - p(\rho_{11} - \rho_{00}), \quad (5a)$$
$$\dot{\rho}_{00} = -\gamma_0(\rho_{00} - n_0^0) + p(\rho_{11} - \rho_{00}), \quad (5b)$$

where $\rho_{11}$ and $\rho_{00}$ are the populations of the $m_s = 0$ and $m_s = +1$ states, respectively, $p$ is the effective MW mixing rate, $\gamma_0$ and $\gamma_1$ are the population relaxation rates, and $n_0^0$ and $n_1^0$ are the equilibrium populations of the $m_s = 0$ and $m_s = +1$ states in the presence of optical excitation. In our modeling we postulate the effective field resulting from beating of the individual MW fields, so that $p(t)$ takes the form:

$$p(t) = \left\{1 + F\frac{\gamma^2}{\gamma^2+(\beta t)^2}\cos[(\omega_0 + \beta t)t + \phi]\right\}p_0. \quad (6)$$

Here, $p_0$ is the mean pump rate, $p_0 = \Omega^2/\Gamma$, $\beta = \partial\Delta_\omega/\partial t$ is the scan (chirp) rate, and F is the contrast factor reflecting the ratio between the pump and probe powers. The Lorentzian term in Eq.(6) reflects the limited temporal response of the populations to the modulation of the total MW field and, consequently, reflects the finite width of the resonances observed as a function of ω$_p$ (Fig. 5). Out of resonance, the dynamics of populations is given by the total MW power (p$_0$), while on-resonance the system is driven by the amplitude modulated field $p(t)$ due to the cosine term. As the time dependence of $p(t)$ becomes slow for nearly equal frequencies of the beating fields, we solve Eqs. (5) in a steady-state approximation, which yields the population difference equal to:

$$\rho_{00}(t) - \rho_{11}(t) = \frac{\gamma}{p(t)+\gamma}(1 - 2n_1^0). \quad (7)$$

In this simple analysis, the initial population $n_1^0$ is responsible for the amplitude of the signal, while the temporal changes of *p(t)* result in the oscillatory behavior of the ODMR signal, which is proportional to the left-hand side of Eq. (7). For the comparable pump and probe powers, the rate *p(t)* oscillates between zero and its maximum value, which for *p>γ* create strong nonlinearity of the ODMR signal. The amplitude of the population oscillations corresponds to the full depth of the ODMR line in a single MW field experiment (black curve in Fig. 3 (a)), rather than that of a pump-probe situation (red curve, Fig. 3 (a)).

To verify the model and get more physical insight, we have simulated the difference of populations for the chirp rates corresponding to the ones in Fig. 6, assuming equilibrium population of $m_s = +1$ state equal $n_1^0 = 0.1$, the Rabi frequency of $\Omega_p/2\pi = 0.4$ MHz and the relaxation rates equal to: $\gamma_0/2\pi = 10$ kHz, $\gamma_1/2\pi = 50$ kHz, and $\Gamma/2\pi = 5.8$ MHz. The initial beating phase was adjusted to match the experimental data. Results of the simulations are shown in Fig. 9. They agree qualitatively with the signals shown in Fig. 6, the main difference being the shape of the envelope, which was taken as Lorentzian in our model, while the experimental data show more pointed envelope shape around zero detuning. We believe the difference is caused by the simplicity of our beating model, like neglecting of nonlinear interaction with the probe field and higher-order mixing terms in Eq. (1).

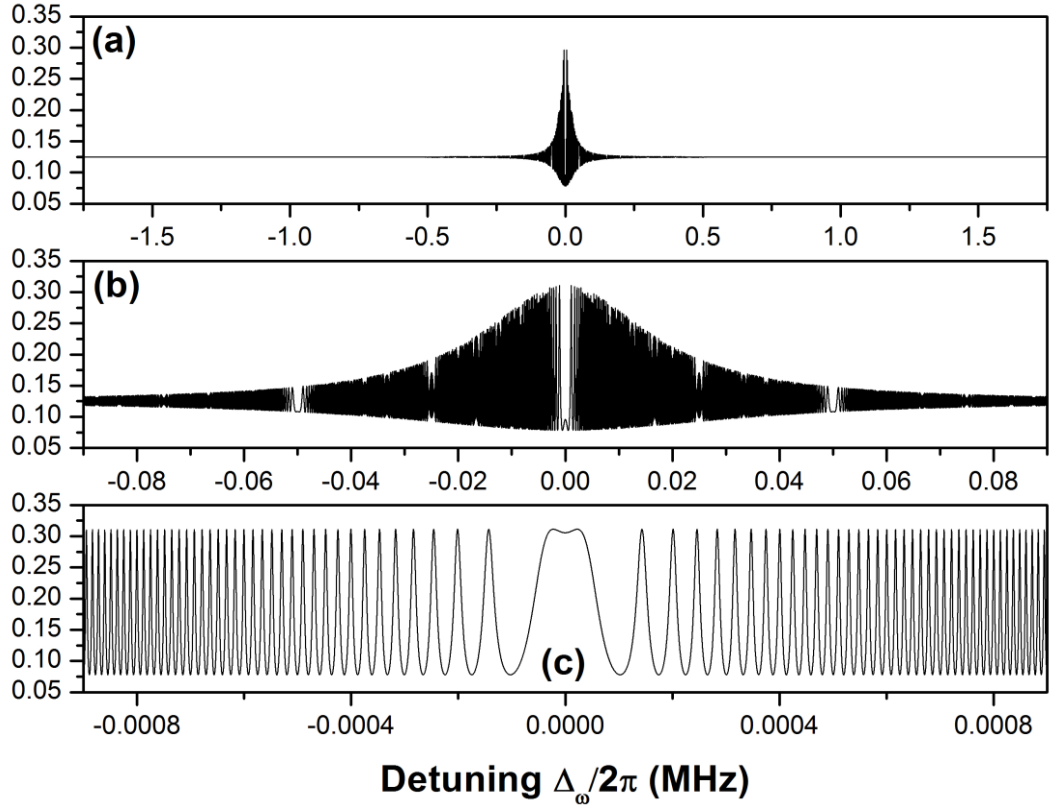

Fig.9. Simulation of the population difference in the system calculated using rate-equations and quasi-stationary effective mixing field with the parameters corresponding to those of Fig.6. (periodic signal drops seen in (a) and (b) are numerical artefacts).

## 7. CONCLUSIONS

We have presented detailed studies of the MW hole-burning in NV⁻ diamond when the pump and probe MW fields are acting on the same transition in the ground state of NV⁻ color center which extend and complement the case described in Ref. [10] where the pumping and probing occurred on distinct transitions. The experimental results, as well as our theoretical modeling, indicate that in the case of the same transition the effect of coherent population oscillation (CPO) strongly affects the observed ODMR spectra.

We have observed the population oscillations in the intensity of fluorescence of the investigated sample in the spectral domain as a function of MW frequency/phase, and in the time domain, as a function of time. The oscillations observed in the time domain occur when the frequencies of two MWs are close to each other and to the frequency of the spin transition. They have a rich harmonic content with the number of harmonics increasing with the MW power. To our knowledge, this is the first direct observation of the CPO in real time and the first observation of CPO effects in NV diamond.

The oscillations can be resolved in a narrow frequency range determined by the relaxation properties of the sample. In that range the population oscillations reach a high amplitude but become sensitive to the relative phase of the MW fields. Consequently, retrieving of the spectral information requires phase averaging. With proper averaging, we have recorded narrow MW resonances in the ODMR spectrum, which have a complex structure composed of three Lorentzian resonances positioned at the pump-field frequency and having their widths and amplitudes dependent on the lifetimes of the levels involved in the transition.

The described oscillations and hole-burning lineshapes in the +/+ configuration are specific for the NV samples: (1) They do not occur when the MW frequencies are detuned from the transitions between the spin states; (2) They occur in time domain as real-time oscillations in a narrow bandwidth and exhibit widths characteristic for the NV sample; (3) They agree well with the theoretical modelling of CPO, verified by several earlier experiments in the optical domain.

These observations give insight into the nonlinear dynamics of the system consisting of an NV[-] sample and microwave field. Specifically, their sensitivity to the relaxation dynamics of the NV sample suggests using the CPO resonances as an alternative method for studying the relaxation processes in NV[-] samples. Careful analysis of the +/+ hole shape yields three relaxation constants $\gamma_0$ and $\gamma_1$, which can be related to the populations of $m_s = 0$ and $m_s = +1$ states, and $\Gamma$, which reflects the decoherence rate of the studied NV[-] system. While $\Gamma$ is directly associated with the transverse relaxation time $\Gamma = 1/T_2$, the constants $\gamma_0$ and $\gamma_1$ determine the longitudinal relaxation rate, $½(\gamma_0+\gamma_1) = 1/T_1$. As compared with other methods, like relaxation in the dark with delayed optical readout [44,45], a possible advantage of CPO would be its specificity to relaxation and dephasing of individual states of the NV sample. CPO method is also free from some assumptions inherent in other methods (see the discussion in Ref. [45]). On the other hand, one has to remember that the measured relaxation constants characterize the system perturbed by the light, and MWs, so that a comparison with other measurements requires extrapolation of the measured results to zero light intensity and MW power. As pointed out in the Appendix, in the case of narrow resonance profiles the fitting procedure may be inaccurate if the individual profiles do not differ much.

In conclusion, we have analyzed the properties of the hole-burning spectroscopy with two MW fields which share the same spin transition. We have found that the hole shape is determined by CPOs and were able to demonstrate directly the population oscillations. Analysis of the MW spectra suggests a practical use of the CPO resonances as an alternative method for measurements of the individual lifetimes in the NV[-] ground state manifold.

**ACKNOWLEDGEMENTS**

This research was supported by the Polish National Science Center (NCN grant 2012/07/B/ST2/00251). DB acknowledges the support by the AFOSR/DARPA QuASAR program, and by DFG through the DIP program (FO 703/2-1). The authors acknowledge stimulating discussion with Arlene Wilson-Gordon, Vladimir Akulin, Moshe Cooper, Ron Folman, Janusz Mlynarczyk, and Harazi Sivan.

# APPENDIX 1: Beat-note of two frequencies vs. amplitude modulation of a single frequency.

In this appendix we discuss the differences between two schemes in which CPO can be observed: the beating of two independent microwave fields and the amplitude modulation of a single-frequency field.

### A) AM modulation

What is typically considered as an amplitude modulated wave can be written in a form:

$$F(t) = A(t)\cos(\omega_c t + \varphi_c) = A_0[1 + m\cos(\omega_M t + \varphi_M)]\cos(\omega_c t + \varphi_c), \quad (A.1)$$

where $A_0$ is the wave's amplitude, $m$ is the modulation index (in the range of 0 to 1), $\omega_c$ and $\omega_M$ are the carrier and modulation frequencies, respectively, while $\varphi_c$ and $\varphi_M$ stand for their corresponding phases. Importantly, the amplitude $A(t)$ is always a non-negative number.

Neglecting the phases, ~~the~~ equation (A.1) can be rewritten to a form:

$$F(t) = A_0 \cos(\omega_c t) + \frac{A_0 m}{2}\cos[(\omega_c + \omega_M)t] + \frac{A_0 m}{2}\cos[(\omega_c - \omega_M)t], \quad (A.2)$$

where one identifies the carrier frequency and two sidebands separated by $2\omega_M$ from each other. In the full modulation case, $m = 1$, the carrier wave carries 2/3, and each sideband only 1/6[th] of the full power.

One could also envisage ~~the sine~~ modulation of the field power rather than amplitude. In this case, the term $A(t)$ should be replaced with $\sqrt{A(t)}$, which again is a non-negative number.

### B) Two-frequency beat-note.

In the case of our experiment the MW field consists of two independent waves $G_1(t)$ and $G_2(t)$ with constant amplitudes:

$$G(t) = G_1(t) + G_2(t) = A_1 \cos(\omega_1 t + \varphi_1) + A_2 \cos(\omega_2 t + \varphi_2). \quad (A.3)$$

We assume $\omega_1 > \omega_2$, neglect the phases again and equalize the amplitudes by setting $A_1 = A_2$. In this case, each wave carries half of the power and eq. (A.3) takes the form ~~of~~:

$$G(t) = 2A_1 \cos(\frac{\omega_1 - \omega_2}{2}t)\cos(\frac{\omega_1 + \omega_2}{2}t), \quad (A\ 4)$$

which resembles eq.(A 1) when we set $\omega_c = \frac{\omega_1 + \omega_2}{2}$ and the beat-note frequency $\omega_{beat} = \frac{\omega_1 - \omega_2}{2}$. The key difference, however, is that now the amplitude, given by the term: $2A_1 \cos(\omega_{beat} t)$, oscillates between $\pm 2A_1$. The negative values of the amplitude can be viewed as a $\pi$ shift of the carrier wave phase, which occurs every half of the modulation period. An important NMR technique exploiting such a phase jumps is the rotary echo [46], where the carrier phase is periodically being shifted by $\pi$.

Let's further rewrite eq. (A 4) using simple trigonometry to the form:

$$G(t) = \pm 2A_1\sqrt{\cos^2(\omega_{beat} t)}\cos(\omega_c t) = \pm 2A_1\sqrt{[1 + \cos(2\omega_{beat}\ t)]/2}\cos(\omega_c t), \quad (A\ 5)$$

where one has to keep the appropriate sign of the square root (which changes every half-period of the modulation). Here one can clearly see the difference between $G(t)$ and $F(t)$–given by eq. (A.1) in a form of a square root of the modulation term and an alternating sign. The former can be achieved by the cosine modulation of the power rather then amplitude of the MW field. The experimental realization of the carrier phase reversal, however, requires more complex modulation schemes, as simple amplitude modulators in the form of rf mixers/attenuators, variable gain amplifiers or choppers (in case of optical fields) generally preserve the carrier phase.

The above reasoning shows that $F$ and $G$ are clearly not equivalent. In our ODMR experiment, however, the dynamics of the system on a timescales much longer than $T_2$ time is governed by the intensity (power) of the MW field, rather than its amplitude. One way of looking at it is that the direction of the oscillating magnetic part of the MW field (which corresponds to the phase of carrier frequency) in a rotating frame is not important. What is important is the field magnitude, which corresponds to the power of MWs and neglects their phase. Coherent population oscillations can be thus observed for low modulation frequency AM and two-frequency beating because the material (NV) system adiabatically follows the power of the MW field. This allows us also to use the rate equations for modelling NV dynamics, as presented in Sec.6 above. For fast modulation frequencies, coherent processes have to be taken into account and the dynamics becomes sensitive to the phase of the MW field. For this reason, the spectral characteristics obtained in Sec. 4 are applicable only to the two-frequency case.

**APPENDIX 2: Resonance Lineshape**

In this Appendix we present the results of fitting the observed structure of the hole burned in the +/+ configuration to Eq. (3). As discussed in Sec. 4, the observed spectrum may consist of three contributions of different amplitudes and widths. Their separation may be difficult if the widths of individual contributions do not differ much. Below, we depict the results of fitting the observed MW pump-probe spectrum (Fig.6) appearing as a broad line with a narrow peak, by a set of two and three Lorentzians. Qualitatively, both fits are satisfactory. Analysis of the fit residuas reveals, however, the observed signal consists of three, rather than two resonances. Such analysis of the lineshape enables determination of the relaxation rates of the NV$^-$ ensemble under various conditions and should be useful for analysis of relaxation phenomena. In the analyzed case, the difference between the spectra containing two and three narrow contributions is small, so the fitting results need be considered carefully. A better statistics and careful analysis of other experimental conditions, like the light intensity, should allow for more reliable data, allowing meaningful comparison with other measurements.

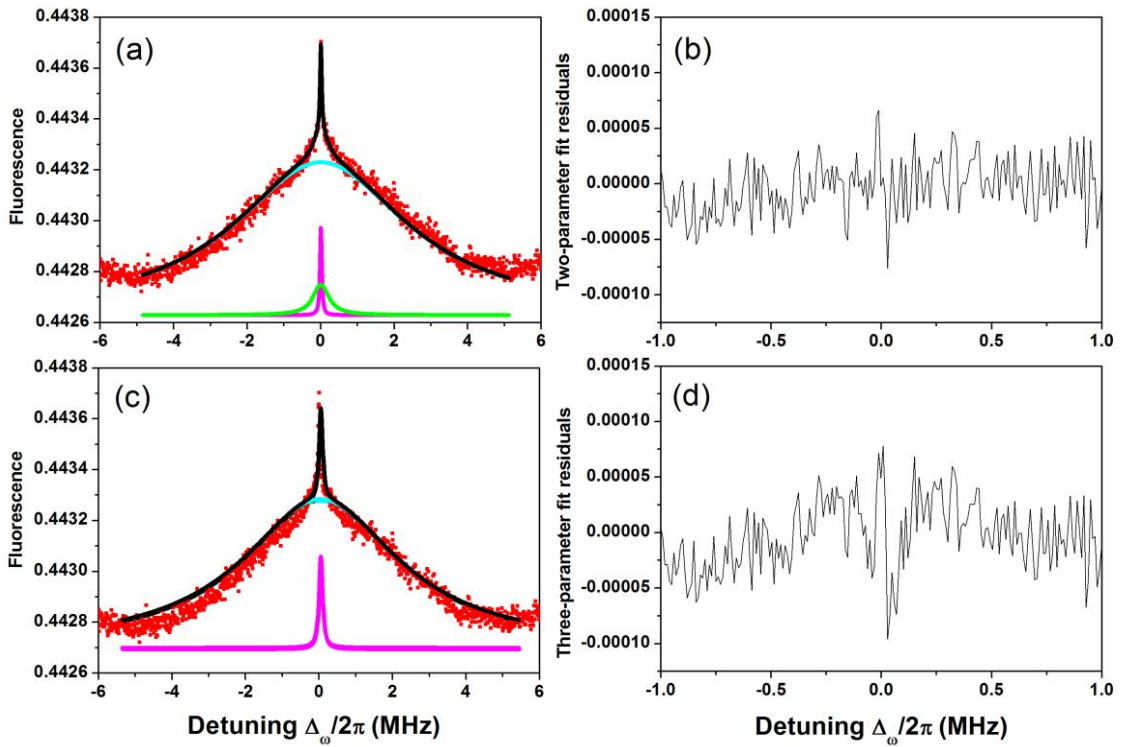

A.1. Fit of the experimental data (red points) from Fig.6 with the sum of three (a) and two (c) Lorentzian profiles (black curve). Also shown are the individual Lorentzian curves. Residuals are shown in (b) and (d), for the (a) and (b) plots, respectively. Note the horizontal scale change.